\documentclass[a4paper]{article}

\usepackage{INTERSPEECH2020}
\usepackage{algorithm}
\usepackage{algorithmic}
\usepackage{multirow}
\usepackage{url}
\usepackage{cite}

\title{Label-Synchronous Speech-to-Text Alignment for ASR \\Using Forward and Backward Transformers}
\name{Yusuke Kida, Tatsuya Komatsu, Masahito Togami}
\address{LINE Corporation}
\email{\{yusuke.kida, komatsu.tatsuya, masahito.togami\}@linecorp.com}

\begin{document}
\maketitle
\begin{abstract}
This paper proposes a novel label-synchronous speech-to-text alignment technique for automatic speech recognition (ASR).
The speech-to-text alignment is a problem of splitting long audio recordings with un-aligned transcripts into utterance-wise pairs of speech and text.
Unlike conventional methods based on frame-synchronous prediction, the proposed method re-defines the speech-to-text alignment as a label-synchronous text mapping problem.
This enables an accurate alignment benefiting from the strong inference ability of the state-of-the-art attention-based encoder-decoder models, which cannot be applied to the conventional methods.
Two different Transformer models named forward Transformer and backward Transformer are respectively used for estimating an initial and final tokens of a given speech segment based on end-of-sentence prediction with teacher-forcing.
Experiments using the corpus of spontaneous Japanese (CSJ) demonstrate that the proposed method provides an accurate utterance-wise alignment, that matches the manually annotated alignment with as few as 0.2\% errors.
It is also confirmed that a Transformer-based hybrid CTC/Attention ASR model using the aligned speech and text pairs as an additional training data reduces character error rates relatively up to 59.0\%, which is significantly better than 39.0\% reduction by a conventional alignment method based on connectionist temporal classification model.
\end{abstract}

%--------------------------------------
\section{Introduction}
%--------------------------------------

End-to-end automatic speech recognition (ASR) has shown remarkable progress with the rapid development of deep neural networks (DNNs).
One of the advantages of the end-to-end approach lies in its simple structure.
It consists of a single neural network model and does not require any other modules such as language models or pronunciation dictionaries, which are essential for conventional hidden Markov model (HMM)-based approach.
On the other hand, the disadvantage of the end-to-end approach is to be ``data-hungry'' due to their number of parameters.
While the simple structure makes it easy to develop ASR models, it requires a huge amount of speech and text pairs to achieve its full performance \cite{amodei2016deep}.
However, the manual transcription of speech is an expensive and time consuming task.

Various methods have been investigated to minimize the costs for collecting the training data.
They are classified into several approaches according to the speech and text requirements.
When only text is available, text-to-speech (TTS)-based augmentation methods can be used to synthesize speech for making the speech and text pairs \cite{Laptev, Wang, Rossenbach}.
Semi-supervised learning can be adopted when only speech is available without transcripts.
The unlabeled speech is firstly recognized by a pre-trained ASR model, and the recognized text is used as the correct transcript \cite{Synnaeve, Karita2, Vesely}.
When the amount of usable audio and text pairs is limited, various data augmentation methods are used such as noise augmentation \cite{Seltzer}, vocal tract length perturbation \cite{Jaitly, kim2019improved}, speed perturbation \cite{Ko} and SpecAugment \cite{Park}.
Alternatively, self-supervised learning has attracted a great deal of attention recently, since it achieved a good performance in many fields using only a small amount of labeled training data \cite{devlin2018bert, chen2020simple, Baevski, Chi, yi2020applying, sadhu2021wav2vec}.

We focus on another realistic situation where a large amount of audio and text pairs are available, but the correct timestamp of each text is not given, i.e., the text is un-aligned.
This situation occurs when we want to use long audio recordings and their transcripts for the model training.
Television broadcasts with subtitles, and publicly available meeting recordings like congress with their minutes are typical examples.
In this situation, various speech-to-text alignment methods can be useful to split the long audio recordings with un-aligned transcripts into utterance-wise pairs of speech and text, so that the transcripts can be used for the model training.
A traditional approach aligns the text by finding an optimal path from the HMM trellis using Viterbi algorithm \cite{Brugnara, Mcauliffe}.
A similar work based on connectionist temporal classification (CTC) model has also been proposed recently \cite{Kurzinger}.
In, \cite{Braunschweiler, Moreno}, the long audio recordings are firstly recognized by a pre-trained ASR model, then the alignment is performed based on text matching between the recognized text and manual transcripts.
These approaches basically align the text in a frame-synchronous way, which requires frame-wise prediction by pre-trained ASR models.
However, the current state-of-the-art ASR is based on label-synchronous token prediction using attention-based encoder-decoder models \cite{Karita, li2020comparison}.
So these conventional frame-synchronous alignment methods cannot benefit from state-of-the-art models such as the Transformer \cite{vaswani2017attention}.

In this paper, we propose a label-synchronous speech-to-text alignment technique for ASR.
Unlike the conventional frame-synchronous approach, the proposed method re-defines the speech-to-text alignment as a label-synchronous text mapping problem.
This enables an accurate alignment benefiting from the strong inference ability of the state-of-the-art attention-based encoder-decoder models.
To our best knowledge, this is the first attempt to apply the attention-based encoder-decoder models to the speech-to-text alignment problem.
Two different Transformer models named forward Transformer and backward Transformer are respectively used for estimating an initial and final tokens of a given speech segment based on end-of-sentence prediction with teacher-forcing.
Experiments using the corpus of spontaneous Japanese (CSJ) demonstrate that the proposed method provides an accurate utterance-wise alignment.
It is also confirmed that the aligned speech and text pairs can be used as additional training data for ASR models.

%The rest of the paper is structured as follows.
%In Section 2 the conventional frame-synchronous alignment is explained in detail.
%Section 3 describes the details of our proposed method.
%The experimental evaluations and results are shown in Section 4 and 5.

%--------------------------------------
%\section{Frame-synchronous alignment}
\section{Alignment approaches}
%--------------------------------------

\subsection{Frame-synchronous alignment}

Frame-synchronous speech-to-text alignment is a problem which maps phonemes or graphemes of the un-aligned transcript to corresponding speech frames.
An example of the frame-synchronous alignment is illustrated in the upper part of Figure~\ref{fig:baseline}.
In this example, the start and end time (frames) are estimated for each phoneme.
Typical frame-synchronous alignment methods require frame-wise prediction by pre-trained ASR models.
Recently, a DNN-based method referred to as CTC-Segmentation \cite{Kurzinger} has been proposed.
CTC-Segmentation generates frame-wise token posteriors using CTC, one of the end-to-end neural network models, and then the alignment is estimated by finding an optimal path from the CTC trellis based on the generated posteriors.
This method has shown superior alignment accuracy compared to the conventional HMM-based methods.
On the other hand, the current state-of-the-art ASR is based on label-synchronous token prediction.
Therefore, these conventional frame-synchronous alignment methods cannot benefit from the state-of-the-art models such as the Transformer.

\subsection{Label-synchronous alignment}

\begin{figure}[tb]
\begin{center}
%\vspace{-10pt}   % 本文と図の間隔微調整用
\includegraphics[width=0.83\columnwidth]{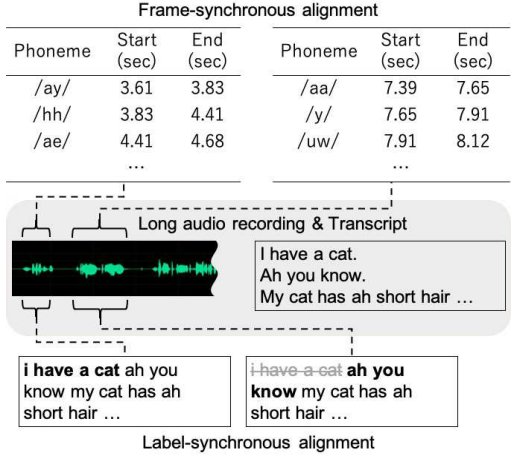}
\vspace{-10pt}     % 図とキャプションの間隔微調整用
\end{center}
\caption{Example of frame-synchronous alignment and label-synchronous alignment.}
\label{fig:baseline}
\vspace{-10pt}   % キャプションと本文の間隔微調整用
\end{figure}

Label-synchronous alignment is newly introduced in this paper.
%Unlike the conventional frame-synchronous alignment methods, the proposed method aligns the text with a label-synchronous approach, which is newly introduced in this paper.
An example of the label-synchronous alignment is shown in the lower part of Figure~\ref{fig:baseline}.
This approach requires the audio signal to be split into short segments in advance, and estimates which part of the transcript corresponds to the given short segment.
The bold text indicates that it is aligned to the segment, and the dashed text indicates that it has already aligned in the past and not subject to the next alignment.
An advantage of this approach is that it can benefit from the strong inference ability of the state-of-the-art attention-based encoder-decoder models.

%--------------------------------------
\section{Proposed method}
%--------------------------------------

\begin{figure}[tb]
\begin{center}
%\vspace{-10pt}   % 本文と図の間隔微調整用
\includegraphics[width=0.84\columnwidth]{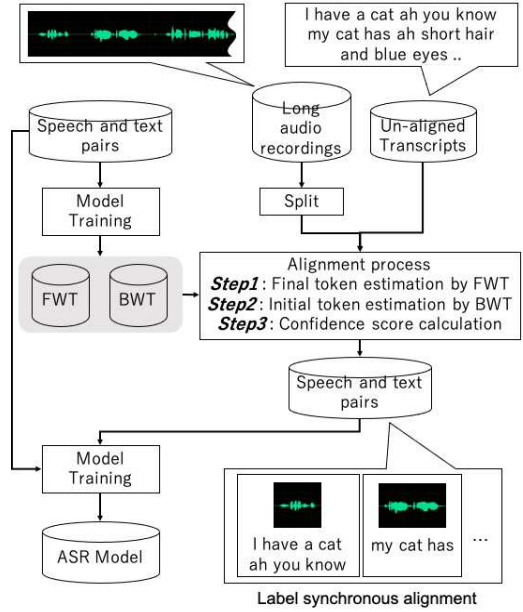}
\vspace{-10pt}     % 図とキャプションの間隔微調整用
\end{center}
\caption{Schematic diagram of proposed speech-to-text alignment method.}
\label{fig:problem}
\vspace{-10pt}   % キャプションと本文の間隔微調整用
\end{figure}

%-----------------------------------------------------------------
\subsection{Overview}
%-----------------------------------------------------------------

A novel speech-to-text alignment method is proposed which is based on the label-synchronous approach.
A schematic diagram of the proposed method is shown in Figure~\ref{fig:problem}.
It is assumed that a long audio recording ${\bf X}$ and un-aligned text transcript ${\bf y}_{1:L}=\{y_1, y_2, \cdots, y_L\}$ are given, where $y_l$ denotes a text token at position $l$.
${\bf X}$ is split into short segments $\{{\bf x}^{(1)}, {\bf x}^{(2)}, \cdots, {\bf x}^{(N)} \}$.
Then an initial token position $l_s^{(n)}$ and final token position $l_e^{(n)}$ of a given segment ${\bf x}^{(n)}$ are respectively estimated from ${\bf y}_{1:L}$, and a token sequence between $l_s^{(n)}$ and $l_e^{(n)}$ is aligned to ${\bf x}^{(n)}$.
The alignment uses two different Transformer models named forward Transformer (FWT) and backward Transformer (BWT), which are trained by already aligned speech and text pairs.
The alignment process is divided into three steps.
(\textbf{Step1}) The final token of a given segment is estimated by FWT finding an end-of-sentence with teacher-forcing.
(\textbf{Step2}) The initial token of a given segment is estimated by BWT reading the tokens backwards from the estimated final token.
(\textbf{Step3}) A confidence score is calculated for the alignment result based on the token posteriors of BWT, and less confident results are rejected.
These steps are repeated until all the segments are processed, and the aligned speech and text pairs are used as an additional training data for the final ASR model.

%--------------------------------------
\subsection{Step1: Final token estimation by FWT}
%--------------------------------------

The final token of a given segment is estimated by FWT finding an end-of-sentence with teacher-forcing.
FWT autoregressively predicts the next token posteriors ${\hat y}_{l}$ using past transcripts ${\bf y}_{l_{start}^{(n)}:l-1}$.
$l_{start}^{(n)}$ indicates a token position where FWT starts to read when $n$-th segment is processed.
FWT reads a token one by one until the model outputs a special token, $\langle \text{eos}\rangle$, indicating an end-of-sentence.
Outputting this token means that there are no more tokens to output in this segment.
Hence, the read token when the model outputs $\langle \text{eos}\rangle$ is considered to be the final token of the segment, and the position index of the final token is estimated as $l_e^{(n)}$.

\begin{figure}[tb]
\begin{center}
%\vspace{-10pt}   % 本文と図の間隔微調整用
\includegraphics[height=45mm]{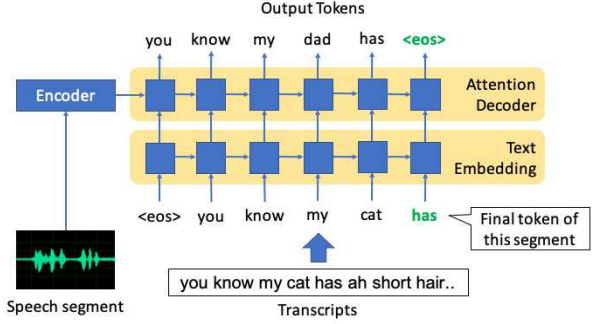}
\vspace{-10pt}     % 図とキャプションの間隔微調整用
\end{center}
\caption{End-token estimation method by FWT.}
\label{fig:fwtransformer}
%\vspace{-10pt}   % キャプションと本文の間隔微調整用
\end{figure}

An example of the decoding by FWT is shown in Figure~\ref{fig:fwtransformer}.
In this example, the model predicts $\langle \text{eos}\rangle$ when the model reads until ``you, know, my, cat, has''.
So ``has'' is regarded as the final token of this segment.
\begin{figure}[tb]
\begin{center}
%\vspace{-10pt}   % 本文と図の間隔微調整用
\includegraphics[height=45mm]{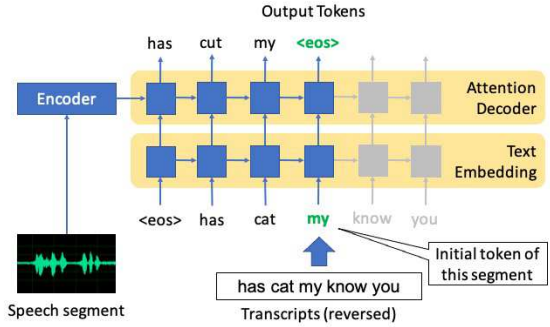}
\vspace{-10pt}     % 図とキャプションの間隔微調整用
\end{center}
\caption{Start-token estimation method by BWT.}
\label{fig:bwtransformer}
%\vspace{-10pt}   % キャプションと本文の間隔微調整用
\end{figure}

%--------------------------------------
\subsection{Step2: Initial token estimation by BWT}
%--------------------------------------

The initial token of the segment is subsequently estimated by BWT, a newly introduced model in this paper.
BWT is a special Transformer model which is trained so that it predicts a token in reverse order.
With the teacher-forcing again, BWT starts to read the token from $l_e^{(n)}$ and goes backward until the model outputs $\langle \text{eos}\rangle$.
Then as well as FWT, the position index of the read token is estimated as $l_s^{(n)}$.

A decoding example for the case of Figure~\ref{fig:fwtransformer} is shown in Figure~\ref{fig:bwtransformer}.
In this example, the model predicts $\langle \text{eos}\rangle$ when the model reads until ``has, cat, my'', and ``my'' is regarded as the initial token of this segment.
Thus, the token sequence between the estimated initial and final tokens ``my, cat, has'' are aligned to the speech segment.

%--------------------------------------
\subsection{Step3: Confidence score calculation}
%--------------------------------------

In the previous steps, only the posteriors for $\langle \text{eos}\rangle$ are considered.
Thus an additional step is conducted to verify the aligned token sequence actually corresponds to the segment.
The verification is based on a confidence score defined as a median value of the posteriors for the aligned token sequence calculated by BWT.
For the above example, if the posteriors of BWT are like \{my, cat, has\} = \{0.75, 0.39, 0.91\}, then the median value is 0.75.
Note that the considered posteriors are not of the estimated token sequence with the highest probability \{my, cut, has\} but the aligned sequence \{my, cat, has\}.
If the confidence score is larger than a pre-defined threshold, the aligned pair is stored.
Otherwise, the aligned pair is discarded, meaning that the aligned speech and text pairs are not used for the ASR model training.

%-----------------------------------------------------------------
\subsection{Overall algorithm}
%-----------------------------------------------------------------

The above steps are repeated until all the speech segments are processed.
The overall algorithm of the proposed method is shown in Algorithm \ref{alg1}.
$Q$ indicates an array containing start positions $l_{start}^{(n)}$ which are defined in 3.2.
${\bf x}^{(n)}$ and $l_{start}^{(n)}$ are firstly loaded (line 4-5).
Then the alignment process based on the above three steps are performed using ${\bf x}^{(n)}$ and $l_{start}^{(n)}$ (line 6-8).
A confidence score $c$ is compared to the threshold $\theta$ (line 9).
If $c$ is larger than $\theta$, the aligned pair of ${\bf x}^{(n)}$ and ${\bf y}_{l_s^{(n)}:l_e^{(n)}}$ are stored (line 10).
Then move to the next segment with resetting $Q$ to $l_e^{(n)}+1$, the next of the estimated final token (line 11-12).
If the confidence is below the threshold, it has possibility that the alignment result is actually correct but falsely rejected.
Another possibility is that this segment only contains noises or laughter that do not need to be aligned with any text tokens.
To deal with both cases, the proposed method appends the next of the estimated final token to current $Q$ (line 14).
\begin{algorithm}[t]
\caption{Proposed method}         
\label{alg1}                          
\begin{algorithmic}[1]
\STATE Segment index: $n=1$
\STATE Array containing starting positions: $Q=\{1\}$
\WHILE{$n \le N$}
\STATE Load $n$-th segment: ${\bf x}^{(n)}$
\FORALL{$l_{start}^{(n)}$ in $Q$}
\STATE \textbf{Step1}: Estimate $l_e^{(n)}$ using FWT
\STATE \textbf{Step2}: Estimate $l_s^{(n)}$ using BWT
\STATE \textbf{Step3}: Calculate confidence score $c$
\IF{$c \geq \theta$}
\STATE Store an aligned pair: \{${\bf x}^{(n)}$, ${\bf y}_{l_s^{(n)}:l_e^{(n)}}$\}
\STATE Reset array: $Q=\{l_e^{(n)}+1\}$
\STATE break
\ELSE
\STATE Append array: $Q=Q.append(l_e^{(n)}+1)$
\ENDIF
\ENDFOR
\STATE $n=n+1$
\ENDWHILE
\end{algorithmic}
\end{algorithm}

%--------------------------------------
\section{Alignment experiment}
%--------------------------------------

A speech-to-text alignment experiment was performed to confirm that the proposed method works as expected.

%-----------------------------------------------------------------
\subsection{Data}
%-----------------------------------------------------------------

The FWT and BWT were trained using all academic presentations (APs) included in the training set of the corpus of spontaneous Japanese (CSJ) \cite{Maekawa}.
The training set contains about 212 hours consisting of 1,006 presentations, and 39 of them were used for the validation set.
Transformer-based hybrid CTC/Attention models were used as FWT and BWT, which were trained using ESPnet \cite{Watanabe}, an open source end-to-end speech processing toolkit.
%Transformer-based hybrid CTC/Attention models \cite{Karita} were trained using ESPnet \cite{Watanabe}, an open source end-to-end speech processing toolkit.
The standard recipe for the CSJ was used for the training, but language models were not used for the decoding.
When training the BWT, the transcripts were input to the model in a reversed order.

The standard three test sets included in the CSJ were used for the evaluation.
Each test set includes 10 speech sessions spoken by a single speaker.
Eval1 and eval2 consisted of APs, and eval3 composed of the simulated public speakings (SPSs).

%-----------------------------------------------------------------
\subsection{Experimental settings}
%-----------------------------------------------------------------

The proposed method assumes that a long audio recording is partitioned into multiple speech segments by some sort of voice activity detection methods \cite{Sohn, hughes2013recurrent, drugman2015voice}.
In this experiment, manually annotated time stamps were instead used for the segmentation, so that the alignment accuracy was evaluated independently of the adopted voice activity detection method.
The evaluation with the voice activity detection will be a future work.
The un-aligned transcripts were simulated by concatenating utterance-wise transcripts together and removing the punctuation marks.
The threshold $\theta$ described in 3.4 was set to 0.7.

The alignment accuracy of the proposed method was evaluated by comparing the aligned text and the original transcript for each speech segment.
The character error rate (CER), which is usually used to evaluate ASR, was also used as the evaluation metrics, since a character is regarded as a token in this experiment.
The error rate was calculated by assuming that the aligned text was the recognized result and the original transcript was the reference.

%-----------------------------------------------------------------
\subsection{Results}
%-----------------------------------------------------------------

The results are shown in Table~\ref{tbl:result-alignment}.
The first row in the table indicates a non-rejected character rate (NRR), the percentage of the aligned characters among all.
The NRRs were over 99\% in all test sets, meaning that almost all of the characters were not rejected by the thresholding scheme described in 3.4.
The CERs of the proposed method shown in the second row were as few as 0.2\%.
These figures were much lower than the CERs obtained when the same FWT model was used for ASR, which are indicated in the third row.
This results indicate that the proposed method worked as expected, and also suggests that the speech and text pairs aligned by the proposed method are expected to be useful as an additional training data for the ASR model training.

\begin{table}
  %\vspace{-10pt}   % 本文とキャプションの間隔微調整用
  \caption{Alignment results of proposed method for CSJ test sets. NRR indicates non-rejected character rate (\%).}
  \label{tbl:result-alignment}
  \begin{center}
  %\vspace{-10pt}   % キャプションと表の間隔微調整用
  \begin{tabular}{c|ccc}\hline
        &  NRR & CER & (Ref) CER by ASR\\ \hline
  Eval1 &  99.7 & 0.5 & 6.0\\
  Eval2 &  99.0 & 0.2 & 4.2\\
  Eval3 &  99.0 & 3.1 & 10.0\\ \hline
  \end{tabular}
  \vspace{-10pt}   % 表と本文の間隔微調整用
  \end{center}
\end{table}

%-----------------------------------------------------------------
\section{ASR experiment}
%-----------------------------------------------------------------

A Japanese ASR experiment was performed to examine if the speech and text pairs aligned by the proposed method actually reduces recognition errors.

%-----------------------------------------------------------------
\subsection{Experimental settings}
%-----------------------------------------------------------------

All SPSs included in the CSJ training set were used in this experiment.
Note that these data were not included in the training set of the FWT and BWT.
They had 329 hours of speech including 1,704 sessions.
The alignment process was performed for each session in a similar way as the above alignment experiment.
The aligned speech and text pairs were augmented to the training set of the FWT and BWT.
A Transformer-based Hybrid CTC/Attention model was adopted for the ASR model, which has the same architecture as the FWT.
The CSJ test sets were also used for the evaluation.

In this experiment, the CTC-Segmentation explained in section 2 was adopted for the baseline of the proposed method.
An implementation in ESPnet \cite{espnet-ctcseg} was used for the experiment.
The same FWT model was used for the CTC-Segmentation, though only a CTC decoder was used for the decoding.
%The FWT model used for the proposed method was also used for the CTC-Segmentation, though only a CTC decoder was used for the decoding.

%-----------------------------------------------------------------
\subsection{Results}
%-----------------------------------------------------------------

The CERs for the CSJ test sets are shown in Table~\ref{tbl:result-asr}.
The first two lines indicate the results of the FWT and BWT.
It is worth noting that the BWT could perform ASR with a slight degradation, even if it generated tokens in reverse order.
``CTC-Seg'' and ``Proposed'' indicate the results of the ASR models whose training data were augmented by CTC-Segmentation and the proposed method, respectively.
The amount of augmented speech was 273 hours for CTC-Segmentation, and 263 hours for the proposed method.
``Full-Set'' indicates the result where the full training set of CSJ including 581 hours was used for the training, which is thought to be a lower bound of CERs in this setup.
From the table, we can see that the both of CTC-Segmentation and the proposed method could reduce CER significantly from ``FWT'' in all test sets.
The reason for the large improvement in eval3 was probably that the SPSs used for the augmentation was not included in the original training set.
Relative error reduction rates against ``FWT'' are shown in the brackets.
The error reduction by the proposed method was up to 59.0\%, showed significantly better than 39.0\% by CTC-Segmentation.
Thus, it was confirmed that the speech and text pairs aligned by the proposed method could be useful as an additional training data for the ASR model training.

\begin{table}
  %\vspace{-10pt}   % 本文とキャプションの間隔微調整用
  \caption{Comparison of CERs between different models and training sets. Figures in brackets indicate relative error reduction rates from FWT.}
  \label{tbl:result-asr}
  \begin{center}
  %\vspace{-10pt}   % キャプションと表の間隔微調整用
  \begin{tabular}{cc|ccc}\hline
  Model          & Training data & Eval1 & Eval2 & Eval3\\ \hline
  FWT            & 212h & 6.0 & 4.2 & 10.0\\
  BWT            & 212h & 7.9 & 5.7 & 14.0\\ \hline
  CTC-Seg        & 485h & 5.4 & 3.8 &  6.1\\
                 &      & (10.0) & (9.5) & (39.0)\\
  Proposed       & 475h & {\bf 5.2} & {\bf 3.6} & {\bf 4.1}\\
                 &      & (13.3) & (14.4) & (59.0)\\ \hline
  Full-Set       & 581h & 4.9 & 3.3 & 3.4  \\
  \hline
  \end{tabular}
  \vspace{-10pt}   % 表と本文の間隔微調整用
  \end{center}
\end{table}

%-----------------------------------------------------------------
\section{Conclusions}
%-----------------------------------------------------------------

In this paper, a novel label-synchronous speech-to-text alignment technique has been proposed for ASR model training.
The proposed method used two different Transformer models for estimating an initial and final tokens of a given speech segment, which enabled an accurate alignment benefiting from the strong inference ability of the state-of-the-art attention-based encoder-decoder models, which could not be applied to the conventional methods.
Experiments demonstrated that the proposed method could provide an accurate utterance-wise alignment, that matched the manually annotated alignment with as few as 0.2\% errors.
It was also confirmed that the ASR model using aligned speech and text pairs as an additional training data showed up to relative 59.0\% improvement in CER, which was significantly better than 39.0\% by a CTC-based conventional frame-wise alignment method.

%-----------------------------------------------------------------
% 参考文献
%-----------------------------------------------------------------
\bibliographystyle{IEEEtran}

\bibliography{bibliography}

\end{document}